\documentclass[aps,prl,twocolumn,groupedaddress,showpacs]{revtex4}

\usepackage{epsf}

\begin{document}

\title{High Precision Measurement of the Superallowed $0^{+} \rightarrow 0^{+} \beta$ Decay of $^{22}$Mg}
\author{J.C. Hardy, V.E. Iacob, M. Sanchez-Vega, R.G. Neilson, A. Azhari, C.A. Gagliardi, V.E. Mayes, X. Tang, L. Trache and R.E. Tribble}
\affiliation{Cyclotron Institute, Texas A\&M University, College Station, Texas 77843}

\date{\today}

\begin{abstract}
The half-life, 3.8755(12) s, and superallowed branching ratio, 0.5315(12), for $^{22}$Mg $\beta$-decay
have been measured with high precision.  The latter depended on $\gamma$-ray
intensities being measured with an HPGe detector calibrated for relative
efficiencies to an unprecedented 0.15\%.  Previous precise measurements of $0^{+} \rightarrow 0^{+}$ transitions have
been restricted to the nine that populate stable daughter nuclei.  No more such cases exist, and
any improvement in a critical CKM unitarity test must depend on precise measurements of more exotic nuclei.  With this branching-ratio measurement, we show those to be possible for $T_z = -1$ parents.  We obtain a corrected ${\cal F}t$-value of 3071(9) s, in good agreement with expectations.
\end{abstract}


\maketitle

Superallowed $0^{+} \rightarrow 0^{+}$ nuclear $\beta$-decay is a sensitive probe of the vector part
of the weak interaction.  Measurement of the $ft$-value for such a transition yields a direct
determination of the vector coupling constant, $G_V$, provided that small radiative corrections are
properly accounted for.  To date, the $ft$-values for nine $0^{+} \rightarrow 0^{+}$ transitions -- 
the decays of $^{10}$C, $^{14}$O, $^{26m}$Al, $^{34}$Cl, $^{38m}$K, $^{42}$Sc, $^{46}$V, $^{50}$Mn and
$^{54}$Co -- have been measured with $\sim 0.1\%$ precision or better, and these results yield fully
consistent values for $G_V$. With $G_V$ thus determined, it is possible to establish a very precise
value for $V_{ud}$, the up-down element of the Cabibbo-Kobayashi-Maskawa (CKM) quark-mixing matrix.  Not
only is this the most precise determination of $V_{ud}$, it is the most precise result for any element
in the CKM matrix.  It also leads to the most demanding test available of CKM unitarity, a fundamental
tenet of the minimal standard model.  Strikingly, the test fails by more than two standard deviations
\cite{TH02,WEIN98}: {\it viz.} $V_{ud}^2 + V_{us}^2 + V_{ub}^2 = 0.9968 \pm 0.0014$.  Since recent results
suggest that the value of $V_{us}$ may need to be revised \cite{Sh02}, it is even more important that the
value of $V_{ud}$ be known as precisely as possible.

Since the uncertainty in $V_{ud}$ is dominated by the uncertainty in the small ($\sim 1\%$) calculated correction terms
that are applied to the data, any improvement in the statistical definition of the unitarity test
must come from improvements in those terms, two of which depend on structure details for the nuclei
involved.  It has recently been argued \cite{TH02} that the best way to validate the structure-dependent
correction terms is to measure additional superallowed transitions specifically selected to cover a wider
range of correction-term magnitudes, demonstrating whether these transitions also produce consistent $G_V$
values.  To this end, we focus on the even-even $T_z = -1$ nuclei with $18 \leq A \leq 42$, selected
because their decays are between nuclei described by the same nuclear-model spaces as those used for some
of the nine currently well known cases.  Since the calculated correction terms are obtained in a completely
consistent way, the tests of these terms will have a direct impact on current results for $V_{ud}$.  The
$T_z = -1$ nuclei, however, are all farther from stability than the currently known cases, have unstable
daughters and exhibit multiple decay branches.  Their measurement presents real experimental challenges.

The $ft$ value that characterizes any $\beta$-transition is determined by three measurable parameters: the
transition energy, $Q_{EC}$, which is used in calculating the statistical rate function $f$; the half-life of
the $\beta$-emitter and the branching ratio for the transition of interest, which together yield the partial
half-life, $t$.  What separates the decay of a $T_z = -1$ superallowed emitter from one with $T_z = 0$ is the
complexity of its decay.  While the latter concentrates $>99\%$ of its total decay strength in the superallowed
branch, the former includes strong Gamow-Teller branches in addition to the superallowed one.  To achieve the
$\sim 0.1\%$ precision required for a meaningful branching-ratio result, all that is required for the latter
case is to measure any non-superallowed branches with modest precision ($\sim 10\%$) and subtract their total
from 100\%; for the former case, the superallowed branch itself must be measured directly and with the full $\sim 0.1\%$
precision.  With the result reported in this letter, $^{22}$Mg becomes the first case of a $T_z = -1$ parent
with an unstable daughter whose branching ratio and half-life have been measured with such high precision.  The
techniques used in this measurement can be applied in future to other similar decays.

A detailed description of these experiments will appear in a later publication \cite{tbp}.  In summary, we produced 3.9-s $^{22}$Mg using a $28A$-MeV $^{23}$Na beam from the Texas A\&M K500 cyclotron to initiate the
$^1$H($^{23}$Na, 2n)$^{22}$Mg reaction on an LN$_2$-cooled hydrogen gas target.  The ejectiles entered the MARS spectrometer
\cite{TR91} where the $^{23}$Na beam was stopped and the fully stripped reaction products were spatially separated
from one another, leaving a $>$$99.6\%$ pure $^{22}$Mg beam at the extraction slits in the MARS focal plane.  This
beam, containing $\sim$$10^4$ atoms/s at $23A$ MeV, then exited the vacuum system through a 50-$\mu$m-thick Kapton window,
passed successively through a 0.3-mm-thick BC-104 scintillator and a stack of aluminum degraders, finally stopping
in the 75-$\mu$m-thick aluminized mylar tape of a tape transport system.  Since the few impurities
remaining in the beam had different ranges from $^{22}$Mg, most were not collected on the tape; residual
collected impurities were found to be substantially less than $0.1$\% of the $^{22}$Mg content.

In a typical measurement, we collected $^{22}$Mg on the tape for 5 s, then interrupted the accelerator
beam in a few $\mu$s by shifting off-resonance the phase of one of the cyclotron dees, and triggered
the tape-transport system to move the sample in 180 ms to a shielded counting station
located 90 cm away.  There, data were recorded for a predetermined counting period while the beam remained
off.  This cycle was clock-controlled and was repeated continuously.  For the branching-ratio measurement,
each counting period was 5 s, during which the sample was positioned between a $70\%$ HPGe $\gamma$-ray
detector and a 1-mm-thick BC404 plastic scintillator that was used to detect $\beta^+$ particles.  The
former was located 15 cm from the sample, while the latter was 3 mm away.  Time-tagged coincidence (or singles)
data were stored event by event.

\begin{figure}[t]
\leavevmode
\epsfysize=7cm
\epsfbox{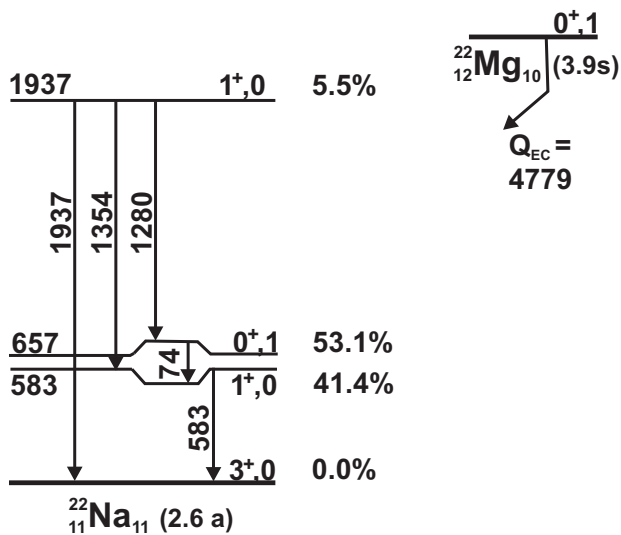}
\caption{Decay scheme for the $\beta$-decay of $^{22}$Mg.}
\label{fig:1}
\end{figure}

For the half-life measurement, a completely different arrangement was used at the shielded counting location.  
In this case, the tape moved the collected sample to the center of a 4$\pi$ proportional gas counter, where
the $\beta^+$ particles were detected and recorded for 80 s, more than 20 half-lives of $^{22}$Mg.  The
counter signals were amplified and sent to a fast discriminator, which
triggered a gate generator.  The gate signals were then multiscaled, with the scaler's channel-advance being
provided by a time-base accurate and stable to 5 ppm.  A separate decay spectrum was thus recorded for each cycle.
The time duration of the non-extendable signal from the gate generator was chosen to be much longer than any
dead-time from the up-stream modules.  This produced a well defined dominant dead-time, which was monitored
continuously during the measurement and later used during analysis to correct the data cycle-by-cycle.  Tests with 1.8-s $^{35}$Ar implanted under identical conditions demonstrated no leakage of activity from the tape.  A similar
system has been used previously and proven effective in the measurement of very precise half-lives \cite{Ko97}.

\begin{figure}[b]
\leavevmode
\epsfxsize=8.7cm
\hspace{-0.3cm}
\epsfbox{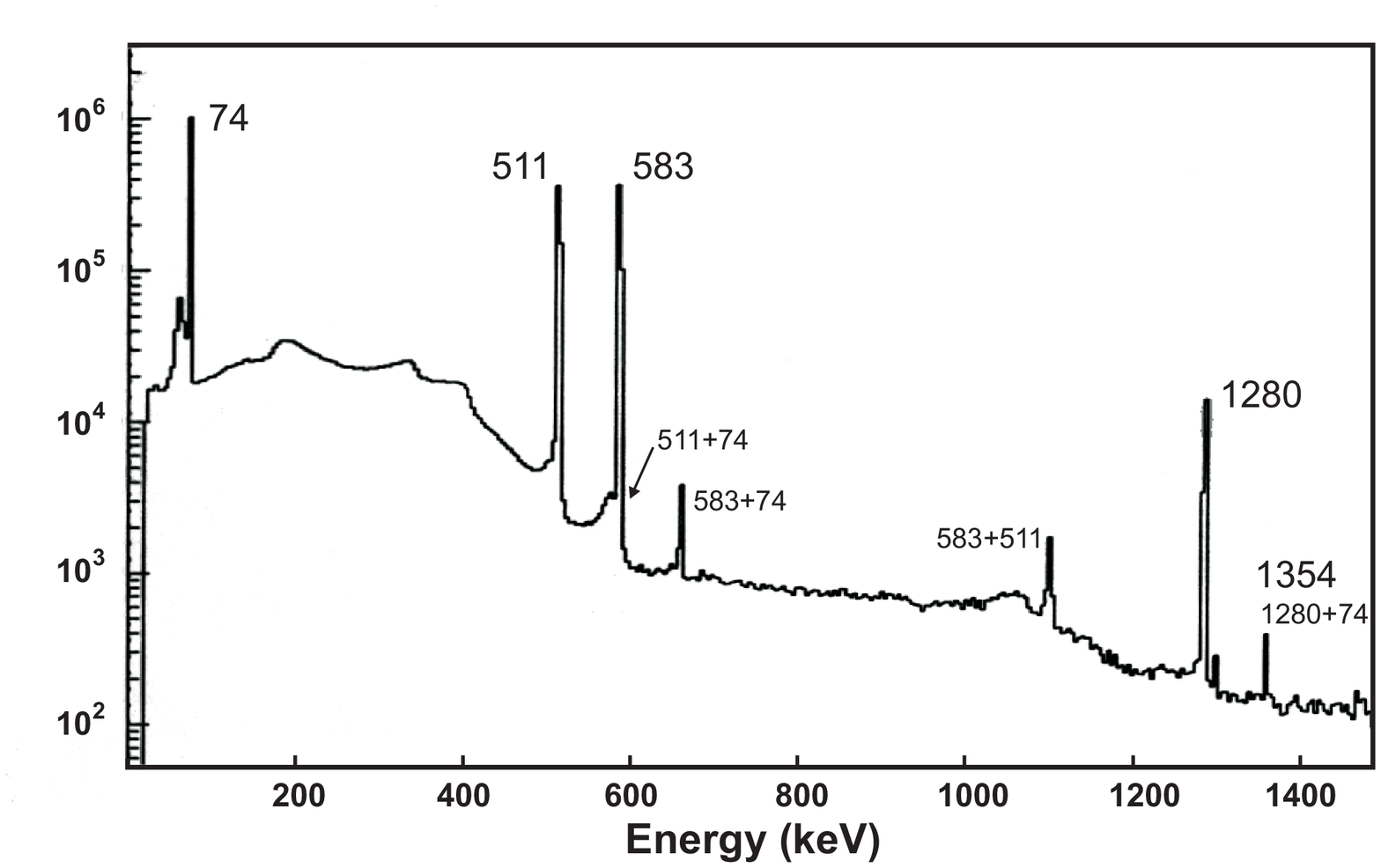}
\caption{Spectrum of $\gamma$-rays observed in coincidence with $\beta$-particles following the decay of $^{22}$Mg.
Primary and coincident-sum peaks are labeled in keV.}
\label{fig:2}
\end{figure}

{\bf Branching ratio:}  The decay scheme of $^{22}$Mg appears in Fig.\ \ref{fig:1}.  Apart from the ground state
of $^{22}$Na -- known \cite{En90} to have ($J^{\pi},T$) = ($3^+,0$) -- it includes the analog ($0^+,1$) state
fed by the superallowed $\beta$ transition, and two ($1^+,0$) states fed by Gamow-Teller branches.  The $\beta$ transition
feeding the ground state must be second-forbidden unique and, being suppressed by some ten orders of magnitude, can
be neglected.  Thus, branching ratios to the excited states can be obtained from the {\it relative} intensities of
$\gamma$-rays observed following the decay of $^{22}$Mg.  Fig.\ \ref{fig:2} shows a portion of the spectrum of these $\gamma$-rays, observed in coincidence with $\beta^+$ particles.

The essence of a precise determination of the superallowed branching ratio is an equally precise measurement of
the relative intensities of the 74 and 583 keV $\gamma$-rays.  To be sure, the intensities of the other three
$\gamma$-rays must be measured carefully too, but much less demanding precision is required.  The primary
measurement was also complicated by three factors peculiar to this decay: 1) HPGe detector signals from a low
energy 74-keV $\gamma$-ray have a wide range of rise times, so special care had to be taken to ensure that none
were missed; 2) the 583 keV state in $^{22}$Na has a half-life of 245 ns, which required us to re-measure this
half-life and to use a 2 $\mu$s $\beta$-$\gamma$ coincidence timing window to minimize the correction for losses from the
583-keV peak; and 3) the 511+74 keV coincident-sum peak is unresolved from the 583 keV peak, a detail that
demanded special attention in analysis.

The fundamental challenge of this measurement, however, was to calibrate the efficiency of an HPGe detector over a
range from 74 to 1937 keV, with the highest possible precision between 74 and 583 keV.  We took data from
thirteen individual sources of ten radionuclides: $^{48}$Cr, $^{60}$Co, $^{88}$Y, $^{108m}$Ag, $^{109}$Cd, $^{120m}$Sb,
$^{133}$Ba, $^{134}$Cs, $^{137}$Cs and $^{180m}$Hf.  Three of these sources we produced ourselves, $^{48}$Cr and
$^{120m}$Sb by using the K500 cyclotron to initiate the $^1$H($^{50}$Cr,p2n)$^{48}$Cr and $^{120}$Sn(p,n)$^{120m}$Sb reactions
respectively, and $^{180m}$Hf by irradiating enriched $^{179}$Hf with thermal neutrons from the Texas A\&M reactor.
Two $^{60}$Co sources were specially prepared by the Physikalisch-Technische Bundasanstalt
\cite {Sc02} with activities certified to 0.06\%.  All other sources were purchased from commercial suppliers.
Sources of greatest importance to a precision calibration are those exhibiting simple $\gamma$-ray cascades uncomplicated
by large conversion-electron components or by any possible $\beta$ side feeding.  Except for the calculable effects of
electron conversion, the intensities of such cascaded $\gamma$-ray transitions are unambiguously equal.  In
particular, the $^{22}$Mg measurement depended most crucially on the following sources (and cascaded $\gamma$-ray energies
in keV): $^{48}$Cr (112.4, 308.3), $^{108m}$Ag (433.9, 614.28, 722.9), $^{120m}$Sb (89.8, 197.3, 1023.1, 1171.3) and
$^{180m}$Hf (215.3, 332.3).  In our analyses of all spectra -- those used for calibration and those taken
in the $^{22}$Mg measurements -- we incorporated corrections for coincidence summing (always $\le 1\%$), and included
effects from angular correlation between each pair of summing $\gamma$-rays.

In addition to acquiring calibration spectra, we also made a number of measurements designed to reveal the physical
dimensions and location of the detector's Ge crystal in its housing.  These measurements included a scan of the side of
the detector with a tightly collimated $^{133}$Ba source, to determine the crystal length; a pair of $^{57}$Co spectra recorded
at 4- and 20-cm source-detector distances, to locate the front surface of the crystal; and an overall x-ray picture of the
crystal in its housing, to establish its exact orientation.  This information was then used as input to Monte Carlo calculations
performed with the electron and photon transport code CYLTRAN \cite{Ha86}.  With
only the detector's two dead-layers as adjustable parameters, we achieved excellent agreement ($\chi^2/N = 0.8$) between the Monte
Carlo efficiency results and our 40 measured data points between 22 and 1836 keV.  With these calculations used to interpolate between
measured calibration efficiencies, we determine a relative efficiency between 74 and 583 keV of 2.876(4), a precision
of $0.15\%$.  Complete calibration details appear elsewhere \cite{Ha02}.

On-line data for the decay of $^{22}$Mg were taken for four widely differing counting rates, thus providing us with the
means to test for any count-rate effects in our results.  No statistically significant differences in the relative
$\gamma$-ray intensities were observed but, to be safe, we eliminated from consideration that quarter of the data with
the highest rate.  From the remaining $\beta$-$\gamma$ coincident events, we next removed the effects ($< 1\%$) of $\gamma$-rays detected in the $\beta$ detector and projected out $\gamma$-ray spectra
that corresponded to different ranges of detected $\beta$ energies.  As expected, no $\beta$-energy dependence was evident
in the ratio of peak areas for the 74- and 583-keV transitions, since both $\gamma$-rays are preceded predominantly by $\beta$-transitions of very similar energy.  However, relative to these two $\gamma$-rays, some $\beta$-energy dependence was evident in the intensities of $\gamma$-rays emanating from the state at 1937 keV, which is fed by a much lower energy $\beta$-decay branch.  Consequently, we used only singles $\gamma$-ray data, which is unaffected by such systematic effects, to obtain the relative intensity of the 1280-keV transition.  Though the effects of room background, negligible in coincidence, had to be incorporated for the singles data, the demands for precision are not nearly so great for the 1280-keV $\gamma$-ray as they are for those at 74 and 583 keV.

Our results for the relative intensities of $\beta$-delayed $\gamma$-rays following the decay of $^{22}$Mg are shown on
the left side of Table~\ref{BR}.  They are generally consistent with, but much more accurate than, previous measurements
\cite{Ha75}; the weak 1354-keV crossover transition has never been identified before.  Taking account of the calculated
conversion coefficient \cite{Ba03}, $\alpha$ = 0.0036, for the 74-keV M1 transition in $^{22}$Na, we used these $\gamma$-ray intensities
to obtain branching ratios, shown on the right side of the table, for the three $^{22}$Mg $\beta$ transitions feeding states in $^{22}$Na. 

\begin{table}
\begin{center}
\caption{Measured relative intensities of $\beta$-delayed $\gamma$-rays, and deduced branching ratios (as \%) for the
$\beta$-decay of $^{22}$Mg. 
\label{BR}}
\vskip 1mm
\begin{ruledtabular}
\begin{tabular}{rll|cl}
& & & \\ [-4mm]
\multicolumn{1}{c}{$E_{\gamma}$}
& \multicolumn{1}{c}{$I_{\gamma}$}
& & \multicolumn{1}{l}{$E_x$($^{22}$Na)}
& \multicolumn{1}{c}{$I_{\beta}$}  \\
& & & & \\ [-4mm]
\hline
& & & & \\ [-3mm]
74 &  ~~58.36(6) & & ~583  & 41.40(13)  \\
583 &  100.00(19) & & ~657  & 53.15(12)  \\
1280 &  ~~~5.40(7) & & 1937  & ~5.45(5)  \\
1354 &  ~~~0.015(3) & & &  \\
1937 &  ~~~0.032(3) & & &  \\         
%& & & &  \\ [-3mm]
\end{tabular}
\end{ruledtabular}
\vspace{-0.5cm}
\end{center}
\end{table}

{\bf Half-life:} We recorded a total of more than 54-million multiscaled decay events from the $4\pi$ $\beta$-detector,
comprising some 3000 collect/count cycles in a sequence of 50 individual measurements, each with a different combination
of detector high-voltage, discriminator and dominant dead-time settings.  The ratio of $^{22}$Mg counts to room background was $\sim 10^{4}$ at the beginning of each counting cycle.  In addition to the 80-s counting periods used for these measurements, we also recorded decay data for a 160-s decay period to identify and characterize any possible impurities of comparable half-life.  Only 22.5-s $^{21}$Na was observed, with an initial activity $1.7\times10^{-4}$ that of $^{22}$Mg, and its effects were incorporated in all analyses.

The data were analyzed with two different fitting procedures to extract the half-life: (i) a
maximum-likelihood fit to the sum of all dead-time corrected decay spectra; and (ii) a global fit of individual cycle
spectra, with a common half-life but with amplitudes and dead-times correctly matched to each cycle.  The second procedure
contains no approximation \cite{Ko97} but both yielded concordant results.  To further consolidate the results, both fitting
procedures and all tests were repeated on a parallel set of Monte-Carlo generated spectra, mimicking the trend of the real
data, but with known half-life and background.  The accurate retrieval of the decay parameters used in the generation of
the artificial data validated the fitting procedures and the final half-life result.

No systematic experimental effects were observed, the results from all 50 individual measurements being statistically
consistent with one another.  A final test for any short-lived impurities was negative as well: we removed all data from the first second of the counting period in each measurement, and re-fitted the remainder; then we repeated the
procedure, removing the first two seconds, three seconds and so on.  Within statistics, the half-life was stable against these changes too.  Our final result for the $^{22}$Mg half-life is 3.8755(12) s.

{\bf $Q_{\bf EC}$ value:}  Since $^{22}$Na is a long-lived nucleus, its mass-excess is well known to $\pm 500$ eV \cite{Au95} and the excitation energy of its $0^+$ excited state, to $\pm 140$ eV \cite{En90}.  As a result, the quality of the $Q_{EC}$-value for the superallowed transition between $^{22}$Mg and $^{22}$Na depends directly on how well the mass-excess of $^{22}$Mg is known.  The most up-to-date published mass tables \cite{Au95} quote a value obtained from two 30-year-old measurements of the $^{24}$Mg(p,t)$^{22}$Mg $Q$-value \cite{Ha74,No74}, neither of which has been corrected for significant changes that have occurred in their calibration-reaction $Q$-values over the years.  In one case \cite{No74}, the method of calibration was complex enough that it is impossible now to update the result; in the other \cite{Ha74}, the measurement was tied directly to the $^{16}$O(p,t)$^{14}$O $Q$-value.  Incorporating the modern value \cite{Au95} for that $Q$-value to update the original $^{24}$Mg(p,t)$^{22}$Mg result \cite{Ha74}, we obtain a mass excess for $^{22}$Mg of $-402(3)$ keV.  This value is supported by a very recent measurement \cite{Bi03} of a resonance in the $^{21}$Na (p,$\gamma$) $^{22}$Mg reaction, from which a mass excess of $-403.2(13)$ keV is inferred.  These values combined yield $Q_{EC} = 4122.1(13)$ keV for the superallowed transition from $^{22}$Mg.

With branching-ratio and half-life results from the present measurements, together with the updated $Q_{EC}$-value just
described, we can now obtain a corrected ${\cal F}t$-value for the superallowed transition from $^{22}$Mg.  We use
the structure-dependent correction term ($0.50\%$) calculated and tabulated in Ref.\,\cite{TH02} inserted into Eq.\,(3) of that reference.  The result is ${\cal F}t = 3071(9)$ s, in excellent agreement with 3072.2(8) s, the average ${\cal F}t$-value for the nine well known cases studied to date.  This agreement provides important confirmation of the structure-dependent calculations \cite{TH02}, which are used in all cases to extract $G_V$.  Furthermore, a mass measurement of $^{22}$Mg with sub-keV precision would reduce the ${\cal F}t$-value uncertainty to $\pm7$ s or $0.2\%$, thus constituting an even more demanding test.  Such precision is possible today with on-line Penning-trap mass spectrometers and, considering the unfortunate history of the $^{22}$Mg mass, we would urge that such a measurement be made.

We have demonstrated for the first time that a high-precision branching ratio can be measured for the superallowed transition
from a $T_z = -1$ parent to an unstable daughter.  Our technique is extendable to other superallowed emitters of the
same type, even those with a strong ground-state branch, since we directly measure the number of collected nuclei in the counting
sample and can thus determine an absolute branching ratio for each observed transition regardless of whether all are observed.

This work was supported by the U.S. Department of Energy under Grant No.\,DE-FG03-93ER40773 and by the Robert A. Welch Foundation.
JCH thanks the Institute for Nuclear Theory at the University of Washington for its hospitality and support
during part of this work.

\end{document}